\documentclass[twocolumn,a4paper,aps,amssymb,dvipdfm]{revtex4}
\usepackage[dvips]{graphicx}
\begin{document}

\title{ The longitudinal spin relaxation of 2d electrons in Si/SiGe
quantum wells in a magnetic field}

\author{Z. Wilamowski$^{1,2}$, W. Jantsch$^1$ }
\address{$^1$Institut f\"{u}r Halbleiterphysik, Johannes Kepler Universit\"{a}t, A-4040 Linz,
Austria.}
\address{$^2$Institute of Physics, Polish Academy of Sciences, Al. Lotnikow 32/46, PL 0668 Warsaw, Poland}

\date{\today}
\begin{abstract}
 The longitudinal spin relaxation time, $T_1$, in a Si/SiGe quantum well is determined
from the saturation of the ESR signal.  We find values of a few
microseconds.  Investigations of $T_1$ as a function of Fermi
energy, concentration of scattering centers and of the momentum
scattering time, $\tau_k$, lead to the conclusion that for high
electron mobility the spin relaxation is ruled by the
Dyakonov-Perel (DP) mechanism while for low mobility the
Elliott-Yaffet mechanism dominates. The DP relaxation is caused by
Bychkov-Rashba coupling. Evaluation of the DP mechanism shows that
$T_1^{-1}$ for high electron mobility can be effectively reduced
by an external magnetic field. The effect of the degenerate
Fermi-Dirac statistics on the DP process is discussed.

\end{abstract}

 \maketitle

\section{Introduction}

 Looking for spin systems suitable for spintronics or
quantum computing devices, the longitudinal spin relaxation time,
$T_1$, is of basic importance.  $T_1$ is ruled by spin-flip
processes and it corresponds to the characteristic spin memory
time. In this paper we investigate $T_1$ in the high mobility 2d
electron gas in a Si/SiGe quantum well, where electrons can be
easily manipulated by illumination with light and by an electric
field \cite{R1,R2}.  We show that for this material system, which
magnetically is one of the cleanest, $T_1$ is of the order of
microseconds whereas the time needed for a spin manipulation by a
microwave magnetic field is by more than two orders of magnitude
shorter.

We also investigate the mechanism for spin relaxation.  Analyzing
the spin relaxation rate as a function of the momentum scattering
rate, $\tau_k^{-1}$, allows to distinguish the Elliott-Yaffet (EY)
mechanism \cite{R3} and the D'yakonov-Perel (DP) mechanism
\cite{R4,R5}.  The EY mechanism describes the probability of a
spin-flip in a momentum scattering event.  This probability is
ruled by spin-orbit coupling and the resulting admixture of a
state with opposite spin projection \cite{R3,R10}:

\begin{equation}\label{1}
(T_1^{-1})_{EY}=\alpha_{EY}\tau_k^{-1}
\end{equation}

The DP relaxation, in contrast, originates from a zero field spin
splitting of the conduction band states \cite{R4,R5,R10}.  For a
Si quantum well the zero field splitting is described by the
Bychkov-Rashba (BR) term \cite{R3,R10}:
\begin{equation}\label{2}
{\cal H}_{BR}=\alpha_{BR}(\bf k\times \bf{\sigma})\cdot
\hat{\bf{e}}_z
\end{equation}
which was shown to exist also for single sided modulation doped Si
quantum wells \cite {R7}.

Here $\sigma$ stands for the vector of a Pauli spin-matrix of a
conduction electron \cite{R8}, $\bf k$ is the k-vector
proportional to the electron momentum, $\hat{\bf{e}}_z$   is a
unit vector perpendicular to the 2d layer and $\alpha _{BR}$ is
the Rashba parameter that depends on the spin-orbit coupling and
details of the interface \cite{R9}.  Momentum scattering causes
also a time dependent modulation of the BR interaction.  As a
consequence, the probability for spin-flips becomes finite.  For
non-quantizing magnetic field, the DP mechanism \cite{R5,R10} is
expected to be proportional to the momentum scattering time:
\begin{equation}\label{3}
(T_1^{-1})_{DP}=\Omega_{BR}^2\tau_k
\end{equation}
where the frequency $\Omega _{BR}$  is proportional to the
k-vector and the BR parameter $\alpha_{BR}$:
\begin{equation}\label{4}
{\bf\Omega_{BR}}^2=\alpha _{BR}{\bf k}/2\hbar
\end{equation}

In this paper we present results obtained from conduction electron
spin resonance (CESR) spectroscopy.  Simultaneous measurements of
CESR, which allows to evaluate $T_1$, and cyclotron resonance
(CR), which allows \cite{R1,R2} to estimate $\tau_k$, permit the
evaluation of $T_1$ as a function of $\tau_k$.  Such data are
obtained from samples with different donor and electron
concentrations.

\section{Samples and experimental results}

Samples were grown by molecular beam epitaxy on 1000 $\Omega$cm
Si(001) substrates, which show complete carrier freeze-out below
30 K.  A 20 nm thick Si channel with tensile in- plane strain was
deposited on a strain-relaxed Si$_{0.75}$Ge$_{0.25}$ buffer layer,
which consists of a 0.5 $\mu$m thick Si$_{0.75}$Ge$_{0.25}$ layer
on top of a 2 $\mu$m thick Si$_{1-x}$Ge$_x$ layer with
compositional grading.  The upper Si$_{0.75}$Ge$_{0.25}$  barrier
was modulation doped with a 12.5 nm thick, nominally undoped
spacer layer, and capped with 5 nm of Si.  Three modulation doped
Si/SiGe structures with different donor concentrations were
examined.  The electron concentration was changed by the light
illumination.

    All measurements were performed with a the standard X-band ESR spectrometer, at a
microwave frequency 9.4 GHz.  The sample were situated in the
center of the rectangular TM$_{201}$ cavity, at the maximum of the
magnetic component of microwave field (which is perpendicular to
the applied magnetic field).  The sample layer was oriented to be
perpendicular to the applied magnetic field (and to electric
component of microwave field).

 \begin{figure}[t]
 \begin{center}
 \includegraphics[width=7cm]{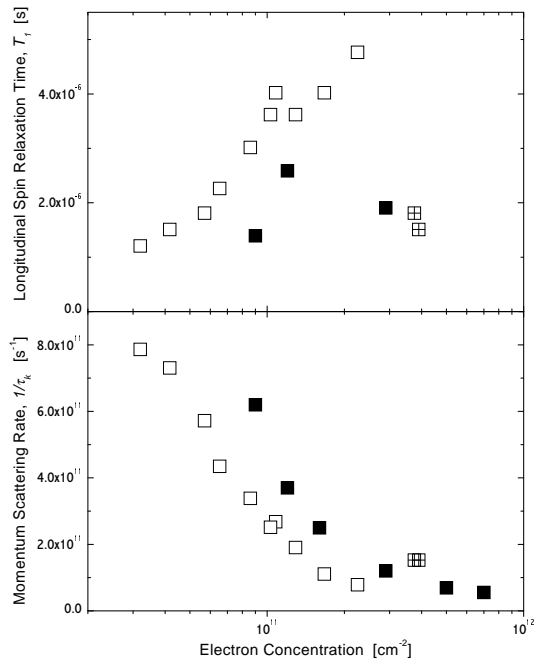}
 \end{center}
 \caption{ (a) Longitudinal spin relaxation time of the
 2d electron gas in a Si/SiGe quantum well
as a function of the sheet electron concentration, $n_s$.  (b)
Concentration dependence of the momentum scattering rate,
$\tau_k^{-1}$, as evaluated from the cyclotron resonance
linewidth. Different symbols stand for samples with different
donor concentrations in the doping layer.}

\label{fig1}
\end{figure}

The spin resonance has an exceedingly narrow linewidth in the
range $3 \div 10 \mu$T.  In spite of the fact that the sample was
situated in the minimum of the electric microwave field and
perpendicular to it, the strong absorption due to CR was well
observed allowing for to monitor the carrier density from the
integral absorption and the momentum scattering rate from the CR
linewidth \cite{R2}.

In Fig. 1 the longitudinal spin relaxation time, $T_1$, and the
momentum relaxation rate, $\tau_k^{-1}$, are plotted as a function
of the electron concentration, $n_s$.  The parameter $T_1$ has
been evaluated from the saturation of the ESR signal amplitude and
the ESR line broadening at high microwave power \cite{R12,R13}.
Estimating the quality factor of the loaded cavity from the
resonance dip width, we obtain the amplitude of the magnetic
component of the microwave field of 1.1 G at a microwave power of
200 mW.  The data for different samples are marked by different
symbols.  The results for the spin relaxation time, $T_1$, vary in
the range of 1 to 5 $\mu$s.  For different samples $T_1$ is
different and it depends on the electron concentration.  The
momentum scattering rate varies with $n_s$ by an order of
magnitude.  The increase of the momentum scattering is related to
the screening breakdown and an increase of the potential
fluctuations at low Fermi energy \cite {R1}.  Samples with a
higher doping level show also a higher $\tau_k^{-1}$.

The dependence of $T_1$  on $n_s$ is governed by the complex
dependence of the relaxation rate on the Fermi k-vector and of the
dependence of $\tau_k^{-1}$ on the electron concentration.  In
order to follow the dependence of the spin relaxation rate,
$T_1^{-1}$, on the momentum relaxation, $\tau_k^{-1}$, our data
are plotted in Figs. 2a and 2b in two different ways.  In Fig. 2a
the spin relaxation rate, $T_1^{-1}$, is given as a function of
the momentum scattering rate, $\tau_k^{-1}$.  In Fig. 2b, the spin
relaxation rate, $T_1^{-1}$, is normalized by the electron
concentration, $n_s$. This normalization allows to account for the
dependence of $T_1^{-1}$ on the BR parameter, $\alpha_{BR}$, and
to study the dependence of the DP rate $T_1^{-1}$ on
$\tau_k^{-1}$.

\begin{figure}[t]
\begin{center}
\includegraphics[width=7cm]{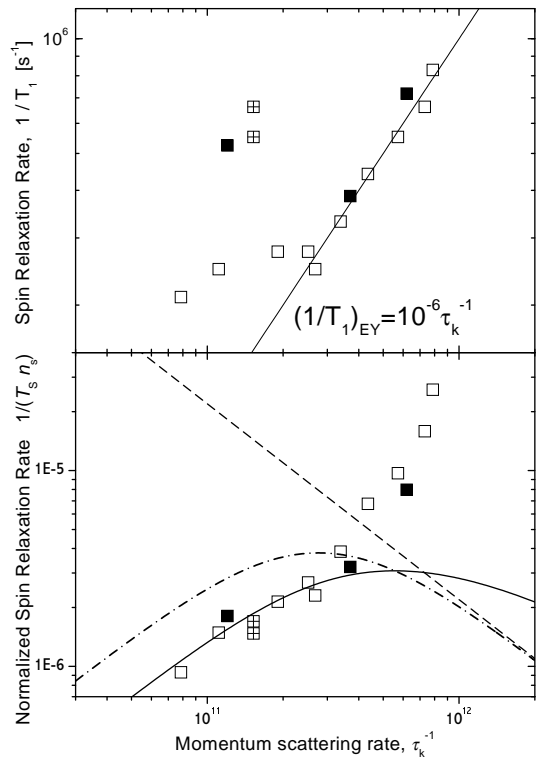}
\end{center}
 \caption{ (a) Spin relaxation rate, $T_1^{-1}$, as a function of
momentum scattering rate,$\tau_k^{-1}$.  It shows that for the
strong scattering the EY process dominates the spin relaxation.
The spin relaxation normalized by the electron concentration,
$n_s$, is shown in Fig (b).  It shows that the spin relaxation for
the high electron mobility, $\tau_k^{-1}<3 \cdot 10^{11}$
s$^{-1}$, is dominated by the DP mechanism.  Solid line: DP
relaxation rate for a BR parameter of $\alpha_{BR}= 1.1 \cdot
10^{11}$ eV·cm \cite{R7} and degenerate statistics (Eqs.(7-8)).
Dash-dotted line: non-degenerate statistics (Eqs. (6-7)).  Dashed
straight line: from Eq. (2) which describes the DP relaxation for
non-degenerate statistics in the absence of an external magnetic
field.} \label{fig2}
\end{figure}

\section{The spin relaxation caused by the Elliott-Yafet mechanism}

Comparison of the data in the two figures demonstrates the
existence of two different ranges with different spin relaxation
behavior.  For high scattering rate, $\tau_k^{-1}>3\cdot10^{11}$
s$^{-1}$, the spin relaxation is simply proportional to the
momentum scattering rate indicating the EY process as the
dominating one.  The EY coefficient is independent both of the
electron concentration and the doping level within the
experimental error.  The solid line in Fig. 2a corresponds to
$\alpha_{EY} = 1.0\cdot 10^{-6}$.

For low momentum scattering rate, $\tau_k^{-1} < 3\cdot10^{11}$
s$^{-1}$, the spin relaxation rate is bigger than expected from
the EY mechanism.  Moreover, $T_1^{-1}$  depends on the electron
concentration in the high electron mobility range.  On the other
hand, the normalized spin relaxation (see Fig. 2b) is
characterized by a systematic dependence, common for all
investigated samples.  As we argue below the observed dependence
in the high mobility range is well described by the DP mechanism.

\section{The D'yakonov-Perel mechanism of the relaxation}

The prediction of the DP scattering rate, as described by Eqs. (3)
and (4) is marked by the dashed line in Fig. 2.  For the BR
parameter we took the value $\alpha_{BR} = 0.55\cdot 10^{-12}$ eV
cm evaluated earlier from the analysis of the linewidth and
g-factor anisotropy in the same samples \cite{R7}.   No
correlation between the dashed line and the experimental data is
recognizable. Eq. (3) stands, however, for the case of a weak
external magnetic field, when the momentum scattering rate is much
smaller as compared to the Zeeman frequency, and it does not
consider cyclotron motion and Landau quantization. In an external
magnetic field, because of the cyclotron motion, the electron
velocity changes its direction all the time.  The time correlation
function of the k-vector, and consequently the correlation
function of the effective BR field seen by an electron, is
describes by:
\begin{equation}\label{5}
\langle {\bf k} \cdot {\bf k}(t)\rangle \propto \langle {\bf
\Omega}_{BR} \cdot {\bf \Omega}_{BR}(t)\rangle=\Omega_{BR}^2
e^{i\omega t-\tau_k^{-1}t}
\end{equation}

The corresponding probabilities of spin-up and down flips are
obtained by the Fourier component of the correlation function
(Eq.(5)) at the Zeeman frequency $\omega_o$:
\begin{equation}\label{6}
W_{\pm}=\frac{\Omega_{BR}^2}{2} \frac{\tau_k}{1+(\omega_c \pm
\omega_o)^2 \tau_k^2}
\end{equation}

The longitudinal spin relaxation time for a single electron is
equal to:
\begin{equation}\label{7}
(T_1^{-1})_{DP}=W_{+}+W_{-}
\end{equation}

Eqs. (6) and (7) describe the DP relaxation in an external
magnetic field.  For a short momentum relaxation time, Eq.(7)
becomes equivalent to Eq.(3).  But for quantizing magnetic field,
where $\omega_c \tau_k>1$ , the DP relaxation rate is expected to
be reduced by the denominator in Eq. (6).  The dependence
corresponding to Eqs. (6-7) is shown in Fig. 2 by the dash-dotted
line.  The reduction of the spin relaxation caused by the external
magnetic field is well visible.  Moreover, for low scattering
rate, $\omega_c \tau_k \gg 1$ , the DP relaxation rate is expected
to be proportional to the momentum scattering.

For an electron gas, in which the final states to which electron
can be scattered, are partially occupied the evaluation of the
mean spin relaxation rate of the whole electron system requires
thermodynamic averaging.  The scattering probability and the
momentum relaxation rate depend on energy.  These quantities are
proportional to the population of empty states:
$\tau_k^{-1}(\varepsilon)=\tau_{ko}^{-1} [1-f_{FD}(\varepsilon)]$
where $\tau_{ko}^{-1}$ is the momentum relaxation rate (as used in
the Boltzmann equation approach) and $f_{FD}(\varepsilon)$ is the
Fermi-Dirac distribution function. For moderate magnetic field,
where $\hbar \omega_c<k_BT<E_F$ the dependence of the BR frequency
on energy can be neglected and the mean value of the transition
probability, $\langle W_{\pm}\rangle$, weighted by the derivative
of the Fermi-Dirac distribution function, is described by:
\begin{equation}\label{8}
\langle W_{\pm}\rangle=\frac{\Omega_{BR}^2}{2}\int
f'_{FD}(\varepsilon)\frac{\tau_k(\varepsilon)}{1+(\omega_c \pm
\omega_o)^2\tau_k^2(\varepsilon)} d\varepsilon
\end{equation}

The solid line in Fig. 2b corresponds to the DP scattering rate as
described by Eqs. (7) and (8).  For the BR parameter again a value
of $0.55\cdot 10^{-12}$ eV cm has been taken.  The theoretical
curve, without any other fitting parameter, fits well to the
experimental points for high electron mobility.  For the highest
mobility the effect of a moderate magnetic field (B=0.34 T) is the
reduction of the DP mechanism by about two orders of magnitude.
In the limit of very high electron mobility, the DP relaxation
rate for degenerate statistics (Eq. (7-8)) is by a factor 2
smaller as compared to the non-degenerate case described by Eqs.
(6-7)

 For the high momentum scattering rate the DP relaxation rate is
expected to tend to the solution for weak magnetic field.  But for
degenerate statistics (solid line), where the final states are
partially occupied, the spin relaxation rate is by a factor 2
bigger as compared to the non-degenerate case (dashed and
dash-dotted lines) for a given momentum relaxation rate.

\section{Conclusions}

In conclusion, we have shown that:
\begin{itemize}
\item the DP mechanism dominates for high mobility structures but the
quantization due to the applied magnetic field leads to a
considerable reduction of the DP relaxation rate.  In Eq. (3) a
reduction factor of about $1+\tau_k^2\omega_c^2$ must be
introduced, where $\omega_c$ is the cyclotron frequency (compare
Eq. (3) and Eqs. (6-7)).  As a consequence, for weak momentum
scattering the reduced DP spin relaxation rate is proportional to
$\tau_k^{-1}$, in contrast to Eq. (3).

\item the value of the BR parameter, $\alpha_{BR}$,
as determined from $T_1$ turns out to be the same within the
experimental accuracy as previously evaluated from the anisotropy
of the CESR linewidth (dephasing time, $T_2$) and the g-factor
\cite{R7}.

\item for low mobility samples the spin relaxation is dominated by the EY mechanism.  We find
an EY coefficient of: $\alpha_{EY}= 1.0 \cdot 10^{-6}$, which is
common for all samples and, for the investigated range of
parameters (the Fermi energy does not exceed 2.5 meV),
$\alpha_{EY}$ does not depend on the electron concentration.

 \end{itemize}

\acknowledgments

We thank F. Sch\"{a}ffler (JKU)
 for generously providing samples and for
helpful discussions.  Work supported within the KBN grant 2 P03B
007 16 in Poland and in Austria by the Fonds zur F\"{o}rderung der
Wissenschaftlichen Forschung, and \"{O}AD, both Vienna.


\begin{thebibliography}{}
\bibitem{R8}
  The coefficient $\alpha_{BR}$, as defined by Equ. (3), is half as big as compared to the case when
  $\sigma$ stands for the spin operator.
\end{thebibliography}
\end{document}